# Measuring Economic Resilience to Natural Disasters with Big Economic Transaction Data

Elena Alfaro Martínez
BBVA Data & Analytics
Madrid, Spain
elena.alfaro@bbvadata.com

Maria Hernández Rubio
BBVA Data & Analytics
Madrid, Spain
maria.hernandezr@bbvadata.com

Roberto Maestre Martinez
BBVA Data & Analytics
Madrid, Spain
roberto.maestre@bbvadata.com

Juan Murillo Arias
BBVA Data & Analytics
Madrid, Spain
juan.murillo.arias@bbvadata.com

Dario Patane
BBVA Data & Analytics
Madrid, Spain
dario.patane@bbvadata.com

Amanda Zerbe
United Nations Global Pulse
New York, NY, USA
amanda@unglobalpulse.org

Robert Kirkpatrick
United Nations Global Pulse
New York, NY, USA
kirkpatrick@unglobalpulse.org

Miguel Luengo-Oroz
United Nations Global Pulse
New York, NY, USA
miguel@unglobalpulse.org

**ABSTRACT**
This research explores the potential to analyze bank card payments and ATM cash withdrawals in order to map and quantify how people are impacted by and recover from natural disasters. Our approach defines a disaster-affected community's economic recovery time as the time needed to return to baseline activity levels in terms of number of bank card payments and ATM cash withdrawals. For Hurricane Odile, which hit the state of Baja California Sur (BCS) in Mexico between 15 and 17 September 2014, we measured and mapped communities' economic recovery time, which ranged from 2 to 40 days in different locations. We found that – among individuals with a bank account – the lower the income level, the shorter the time needed for economic activity to return to normal levels. Gender differences in recovery times were also detected and quantified. In addition, our approach evaluated how communities prepared for the disaster by quantifying expenditure growth in food or gasoline before the hurricane struck. We believe this approach opens a new frontier in measuring the economic impact of disasters with high temporal and spatial resolution, and in understanding how populations bounce back and adapt.

## 1. INTRODUCTION

As factors such as climate change and geopolitical turmoil continue to threaten global efforts to achieve sustainable development, resilience has emerged as a key concept for policymakers. Resilience is the capacity of individuals, communities and systems to absorb and adapt to stress and shocks [1]. It prevents external pressures and shocks from having long-lasting, adverse consequences for development, and ensures that individuals, households, and communities do not fall below a normatively defined level for a given developmental outcome (e.g., food security, poverty level, or wellbeing). Resilience is therefore intimately related to other variables – for example, food security can be viewed as a function of vulnerability, the type and magnitude of shocks, and the resilience of populations.

Quantitative frameworks for measuring resilience [2,3] must begin to take advantage of new opportunities made possible by the ongoing data revolution [4]. There is an urgent need to explore the role of new data sources for quantifying resilience: in particular, certain types of big data such as mobile [5], social media [6] or postal [7] data might afford a more real-time, fine-grained understanding of the resilience of populations confronting various risks [8] that are managed through a combination of household-level, community-level and public measures. Household-level behaviors will be informed by expectations about the nature of the impact and the assistance potentially available (the latter may be particularly important for the most vulnerable and for those anticipating more severe impacts). How might a dense and dynamic data source such as credit card transactions and cash withdrawals serve as a proxy of relevant behavioral changes before, during and after natural disasters?

Commerce and development are closely linked. Payments are proxies for some human interactions, and the current ability to gather and read electronic payment data provides a valuable lens to understand human settlements' activity, societal rhythms and needs, and the influence of external factors on well-being. Recent research conducted by financial institutions has shown that transaction records can be used to detect differences in behavioral patterns related to the socio-economical profile of the cardholder [9], and that aggregate measures of transactions can be used to estimate the tourist attraction capacity of a city [10] or the impact of an event. An interesting recent application of the analysis of this data is the possibility of assessing foreign credit card and mobile phone usage alongside tourism dynamics based on real data, which could be used to give recommendations to the hotel industry [11] or to complement official statistics [12]. Open innovation challenges [13, 14] that have leveraged this type of data have proven to be an exceptional mechanism to foster data use for multiple purposes among developers and entrepreneurs.

Such challenges recognize the wide applicability of this information: transaction data has the potential to provide an accurate view of informative attributes such as the commercial fabric of a territory (density, diversity, evolutions), flows of people and money (at different time scales – hourly, daily, weekly, or seasonally), and areas and sectors of interest for different social segments (residents vs. visitors, young people vs. old, men v. women) [15].

This study aims to derive quantitative proxy indicators of the economic impact of a disaster and the market resilience of the affected population. To that end, the recovery from a shock was measured and mapped for different population subgroups by analyzing the time it took for spending and consumption behaviours to stabilize back to their baseline activity levels.

## 2. MAPPING TRANSACTION PATTERNS BEFORE, DURING AND AFTER A HURRICANE

This research uses transaction data to understand how people in the Mexican state of BCS behaved before, during and after the impact of Hurricane Odile in September 2014. The first major hurricane to hit BCS in 25 years, and tied with Hurricane Olivia (1967) as the most destructive tropical cyclone to strike the region on record, Odile made landfall just east of Cabo San Lucas as a large category 3 hurricane, with windspeed reaching approximately 110kt. 11 direct deaths have been attributed to Odile (8 of which occurred in BCS) and 135 injuries were also reported in BCS. The damages resulted largely from power outages, as Odile took out approximately 550 high-tension transmission towers and 3,400 distribution posts: 90% of the BCS population (over 239,000 people) suffered power outages during the storm. Odile caused enormous damage throughout the coast, leaving thousands homeless in La Paz and communities in cities such as Todos Santos, Pescadero, and Sierra de Laguna having lost virtually all of their possessions. Flooding inhibited transportation and destroyed infrastructure, and considerable damage to the Cabo San Lucas International Airport halted flights (excepting humanitarian aid and relief) for about a month after the disaster. Air travel restrictions left 3,000-4,000 individuals stranded. The economic losses caused by Odile totaled approximately USD $1 billion [16].

In Mexico, 50% of the population has a bank account [17]. The data analyzed in this study represents approximately 30% of all bank account holders in Mexico. This study analyzed data produced by more than 100,000 active clients in BCS (out of an estimated population of 637,000) who produced a total of around 25,000 transactions per day, including point of sale (POS) payments (when someone uses a credit card to pay in a commercial establishment) and ATM cash withdrawals.

To preserve cardholders' and merchants' privacy, no personal identifiable information was part of this study. Data was anonymised and aggregated in compliance with national laws and regulations; individuals cannot be re-identified from any of the results obtained from this study. Despite this robust degree of anonymisation, the raw anonymous registers originally used to produced these aggregated datasets contain rich information, including the transaction itself (timestamp of the purchase, amount spent, credit or debit payment), the store where transactions were made (location, commercial category) and customer demographics (gender, age, home postal code, purchasing power). Each transaction was processed as a unique data point, and no longitudinal data associating multiple data points to a single person was used in the study.

For this study, data has been aggregated into three categories, with volumes of transactions homogeneously distributed throughout BCS: low ($I<$ 120,000MXN/year), medium (120,000MXN/year$<I<$ 250,000MXN/year) and high ($I>$250,000MXN/y). For reference, the mean income in Mexico is around 160,000MXN/year [18]. Based on national income levels published in official statistics in 2014 [19], it can be estimated that the first income group corresponds to 50% of the overall population, the second group to 30%, and the third group to the top 20%. Proportions of men and women were roughly equal for all income groups.

### 2.1. BUILDING BASELINE MODELS AND QUANTIFYING RECOVERY TIMES AND RELATIVE IMPACT

The impact of Odile is highly apparent in the daily time series (Fig. 1B) of total number of transactions in BCS. On the day of the hurricane's arrival, the number of transactions drops precipitously – a phenomenon probably amplified by the associated blackout – which is then followed by recovery until the affected community resumes its previous volume of transactions.

Examining the effect of an event or intervention on a time series can be done through a "synthetic control": a time series that (a) is predictive of the area under examination and (b) is unaffected by the event or intervention, and therefore shows what would have occurred in the absence of that event or intervention. In order to quantify the impact of Odile in BCS, a "normality model" was built using the time series relative to other Mexican regions that were not affected by the hurricane. The "normality model" is based on a Bayesian structural time-series model, and has been implemented using the R library "Causal Impact" [20]. This model made it possible to compare BCS transaction activity to the activity recorded in other regions, in order to estimate what would have happened had the hurricane not occurred. Model over-fitting was controlled by considering only the regions with the highest correlations with BCS.

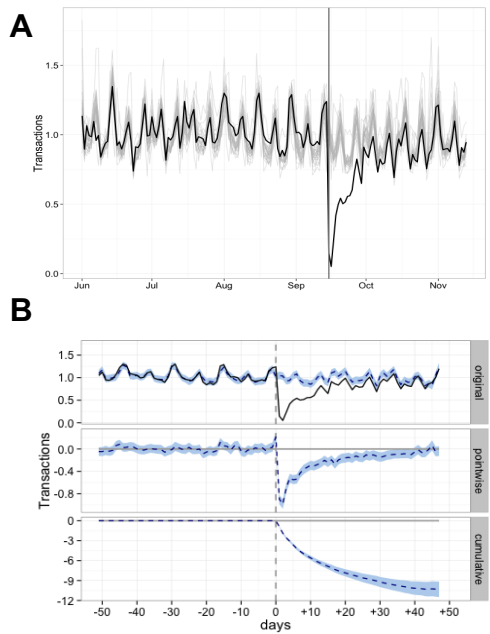
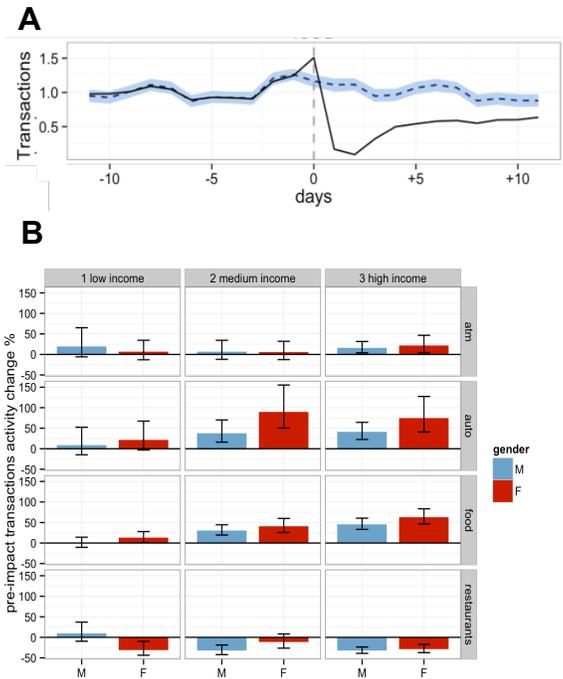

*Fig. 1. (A) Transaction time series for each region in Mexico normalized with respect to their medians. The thick black line is the BCS time series: the negative peak corresponds to the hurricane's landfall. The pale lines represent time series from other regions in Mexico. (B) Upper panel: transaction time series for BCS (black line) and its "normality" model (dashed line), built using the time series for other regions during the pre-impact period. Day 0 corresponds to 2014-09-14, the day before the hurricane made landfall. Middle panel: relative daily impact with respect to the normality model. Lower panel: cumulative impact on transactions in BCS over time.*

### 2.2. DISASTER PREPAREDNESS

The total number of card transactions in BCS was divided roughly equally into ATM and POS transactions. POS expenditures could be further segmented by category. For reference, food accounted for more than 30% of transactions, automobile (including gasoline) for 10%, and bars and restaurants for 10%. The day before the hurricane hit, the spending pattern shifted (Fig. 2): transactions in categories such as food or gas increased (+20%), suggesting that people were stocking up on food and fuel, while the number of transactions in less essential categories dropped (-10%).

The mean amount spent per transaction did not exhibit a parallel shift: therefore, total expenditure increased in the same proportion as the number of transactions. This increase in transactions thus implies that more people were preparing for the event. Analyzing the number of transactions by income level revealed that while all income groups started to prepare at the same time, people with lower income levels spent less than the other groups, in proportion to their normal expenditures, to prepare for the hurricane's arrival. This behavior may have been prompted by a variety of factors: (a) lack of information (a hurricane warning was issued only two days prior to its landfall, due to a change in course as it neared the coast, and followed by the state of emergency the day before landfall); or (b) lack of (or fear of using) personal savings.

*Fig. 2. (A) The number of transactions in the food category scaled with respect to the yearly median (black), and its expected value according to the normality model (blue dashed line). Transactions increased just before the arrival of Odile, reaching a maximum the day before the hurricane made landfall (day 0). (B) Difference between actual volume of transactions and expected normal spending patterns on the day before the event, segmented by income level and gender, for ATM, auto, food, and bars and restaurants categories.*

In terms of gender breakdown, the increase in women's transactions – relative to their expected transaction levels – was roughly double that of men, for all categories except ATM withdrawals in low-income populations

### 2.3. IMPACT AND RECOVERY PER LOCATION

By comparing the normality model to what actually happened, it is possible to estimate both *recovery time* and *relative impact*. We define "recovery time" as the time it takes for the activity to recover to 90% of its "normality" baseline, while the "relative impact" is the cumulative effect (i.e., the total number of predicted transactions that did not occur because of the event during the analyzed period) divided by expected transaction levels.

The normality model was built with transactional data aggregated at the zip code level. Of the 341 zip codes existing in BCS, 131 had sufficient transactional activity to be taken into account in the analysis (determined by setting a minimum number of commercial premises and minimum level of cardholder activity throughout the year, not only during September, 2014). These 131 spatial units were clustered, employing a k-means algorithm over their respective centroids' coordinates, without weighting this coordinates by transactional activity. Therefore, the resulting 8 clusters are a representation of human settlements groups that can automatically be distinguished in BCS.

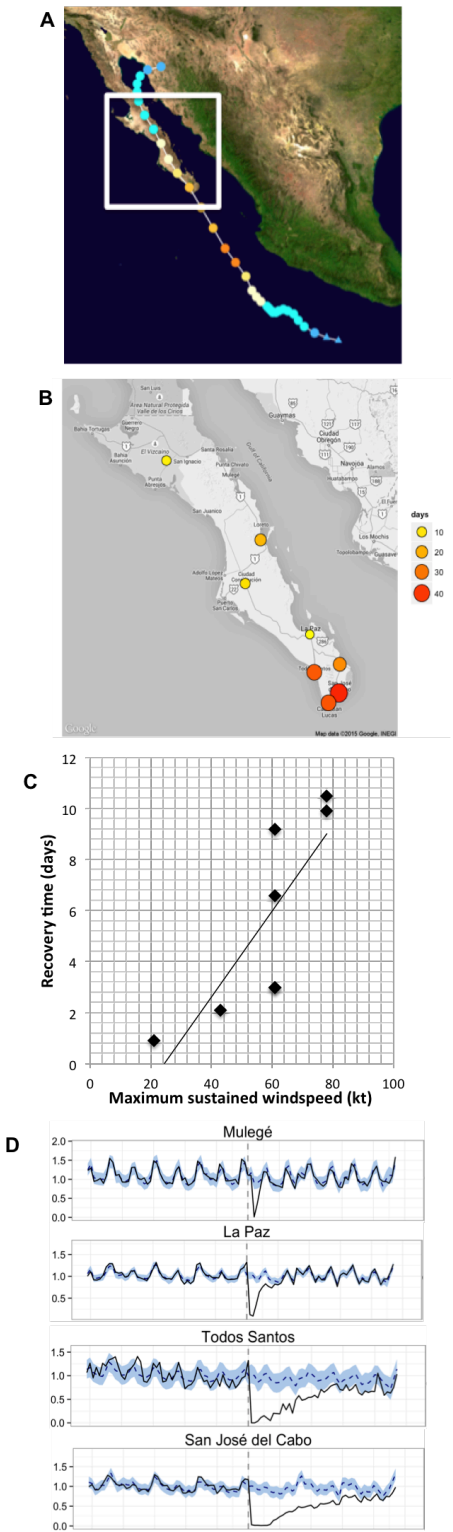

Overall, the average recovery time in BCS is observed to be around 2 weeks. When data is analyzed by location (Fig. 3), however, recovery times vary from 2 days – possibly due to businesses temporarily closing, resulting in an inability to spend – to more than 1 month for the towns located on the south coast where the hurricane struck with its highest intensity. The recovery time calculated was also compared with maximum sustained surface wind speed measurements at nearby weather stations (reported by the US National Hurricane Center) [16]. Higher maximum sustained surface wind speed [1] corresponded with slower recovery times (Fig. 3), a pattern likely due to more damage occurring in these areas.

## 2.4. IMPACT AND RECOVERY PER TYPE OF TRANSACTION

ATM cash withdrawals and retail POS transactions likely represent two distinct preparation strategies for the hurricane. Cash is liquid and thus versatile, making it useful even for evacuees. In contrast, goods are illiquid; an evacuee would have little reason to purchase and transport a large volume of physical goods to a safe location where such goods could also be purchased. For those who remain to ride out the storm, however, goods purchased before the event may be more beneficial than cash, if shops close or run out of stock. These goods are also crucial to meet essential needs in the immediate aftermath of the event. Accordingly, the study also investigated whether households tried to increase holdings of either cash or supplies before or after the disaster.

There was no major alteration of the distribution between cash withdrawals and card purchases patterns in the days prior to landfall, but following landfall, the ratio changed in favour of cash withdrawals (Fig. 4). One potential explanation for this shift may be that in the aftermath of the hurricane, shops closed, reducing opportunities for credit purchases.

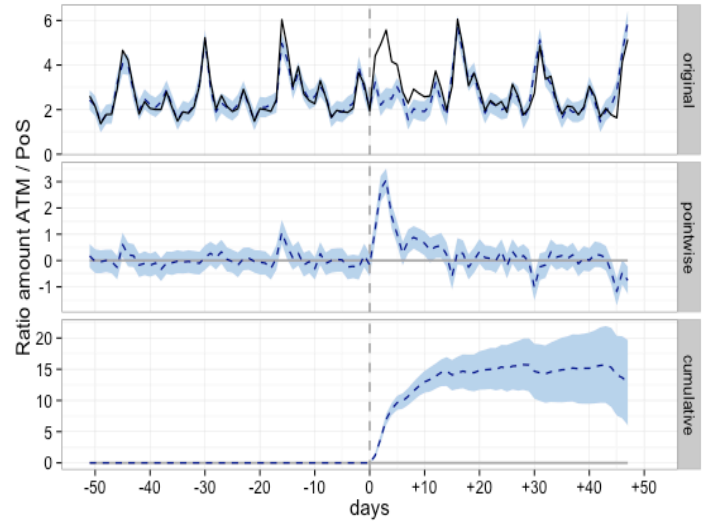

*Fig. 3. (A) Odile's path from south to north. (B) Recovery time for transactions in different locations. (C) Maximum sustained windspeed as measured at nearby weather stations [16] versus economic recovery time at closest population center ($R^2=0.66$). (D) Transaction time series for four different locations, from the northernmost (Mulegé) to the southernmost (San José del Cabo): the black lines are the measured time series while the dashed lines correspond to the normality model.*

*Fig. 4. Evolution of the ratio of ATM to POS transaction in the days around the event, in comparison to the values predicted by the normality model.*

---

[1] Maximum sustained wind speed is defined as the highest 1-minute surface winds occurring in the system circulation, estimated or observed to occur at 10 m in an unobstructed area.

## 2.5. IMPACT AND RECOVERY ACCORDING TO INCOME LEVEL

Analysis of recovery time by income level reveals a tendency for people with lower income to recover faster. This pattern is more evident for ATM transactions (Fig. 5), and may be related to the lower financial capacity of low-income people (which is also reflected in the lower level of pre-disaster preparation activity observed): because low-income people have a reduced capacity to stockpile in preparation for the event, their purchasing levels may need to return to baseline more quickly than those who prepared more extensively. Alternatively, because lower-income people have lower baseline spending levels overall, the time needed to return to baseline transaction levels may be shorter. The shorter recovery time for ATM transactions compared to transactions in commercial premises might be due to the time needed by commercial establishments to reopen (as in some cases electricity supply was cut for several days). Women consistently take slightly longer to return to baseline levels than men.

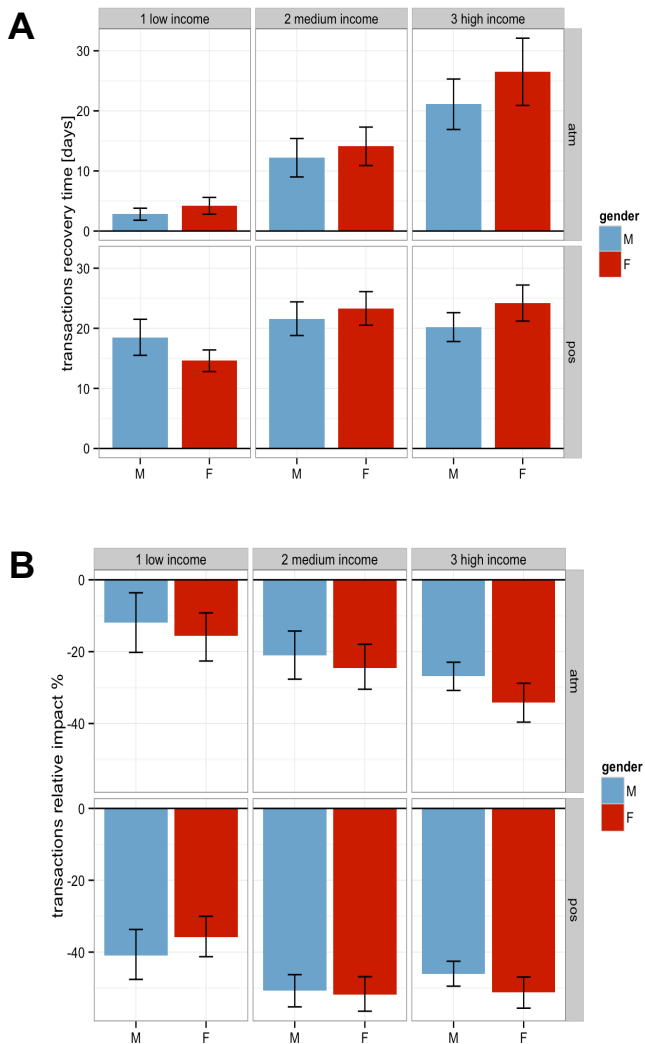

*Fig. 5. (A) Recovery time estimation for ATM and POS activities for low, medium, and high-income individuals. (B) Relative loss estimation (% of cumulative number of transactions) during the four weeks that followed the event. Different POS categories have similar recovery time and impact.*

## 3. RESULTS

The economic impact of, and communities' recovery from, Hurricane Odile in the Mexican State of Baja California Sur was measured and mapped at high frequency and high resolution, using financial transaction data.

Monetary investment in disaster preparedness at the household level was estimated for the day immediately before the hurricane struck, by quantifying increases in expenditure in categories such as food or gasoline (total increase +50%). The mean amount per transaction did not change, implying that this increase was due to more people preparing for the event. In addition, the higher the income group, the more people spent overall – in proportion to their income – in the days before the hurricane struck.

After Hurricane Odile made landfall, economic activity decreased, and average measured recovery time was approximately 2 weeks for point-of-sale transactions and 1 week for ATM cash withdrawals.

In the month after Odile made landfall, 30% fewer POS transactions than expected took place in BCS, and 12% less cash was withdrawn.

Data was also disaggregated by location inside BCS. The highest impact was evident in the southernmost part of the peninsula where the hurricane struck with the highest intensity (with recovery times of up to 40 days for POS transactions).

Analysis of transactions by cardholder income group revealed that the lower the income group, the shorter the recovery time – especially for ATM withdrawals, which returned to normal levels more quickly for the low-income group (2-3 days) than for the medium and high-income groups (more than 10 days).

Women increased their expenditures in preparation for Odile twice as much as men. Recovery times for women are consistently longer than for men.

## 4. DISCUSSION AND FUTURE WORK

This study assessed different impacts and recovery scenarios across close geographies, suggesting the potential of using such information to estimate economic loss at the local level in the wake of a natural disaster, and to inform responses that are highly targeted to the most affected communities. Access to a stream of objective, real-time information on economic recovery could be used to design feedback loops into reconstruction programs and policies. Insights from transaction data could also be incorporated into current mechanisms for estimating the economic losses caused by disasters. For instance, such data could be factored into small business insurance or recovery packages, as low revenues during the recovery period could make it harder for businesses to resume normal operations. Further investigation is warranted into the reasons why low-income populations prepare less for disasters (e.g., lack of risk awareness vs. lack of savings or credit). A potential follow-up study could also consider the likelihood of long-term negative impacts on low-income populations, and/or how coping mechanisms relate to purchasing patterns.

Critically from a privacy protection standpoint, the proposed methodology uses aggregates and statistics derived from individual transactions, which cannot be used to re-identify any particular user. Even with access to the raw data used to produce the measurements applied in this research, it is not possible to link two different transactions to a single client, significantly reducing potential privacy risks [21].

The quantitative information obtained here on disaster preparedness measures undertaken by the population suggests a potential role for proactive, targeted distribution of supplies or cash transfers to the most vulnerable, at-risk populations. The data, when disaggregated by gender, reveal not only the extent to which women prepare households for such disasters, but also that women exhibit longer recovery times than men following a disaster. Such quantitative measures could be used as proxies to infer economic gender gaps. Further investigation would also be required to leverage the potential of continuous transaction monitoring (1) to manage inventory in stores as people prepare, and avoid depletion of essential items; (2) to assess the effectiveness of recovery efforts; (3) to improve targeting of reconstruction aid after a disaster; (4) to proactively target social protection measures, such as insurance for the poor; and/or (5) to develop models that could be used to simulate the economic impact of imminent or potential disasters.

The methodology developed in this study has the potential to be applied in other scenarios to quantitatively estimate the economic impact of a natural disaster. However, further research is needed to understand the similarities and differences of different disasters through the lens of transaction data. Is it possible to model with a similar mathematical function the impact and recovery of different disasters in different places, or in different places for the same disaster? If that is the case, could changes in the parameters needed to fit the model be used to evaluate the effectiveness of disaster preparedness investments, or to quantify the resilience of populations? Ideally, a resilient system would not only adapt from shocks and return to baseline activity levels, but even become stronger and more resilient following the event (i.e., in some cases, transaction levels might even achieve a higher level than before).

Potential next steps could also involve developing the tools and approaches needed to transition from case studies to operational use on-site during disasters, and exploring the potential of such tools and insights to inform humanitarian aid or relief efforts.

## 5. ACKNOWLEDGEMENTS

We would like to acknowledge Shantanu Markeerje (UNDP), Eva Kaplan (UNICEF), and Joanna Syroka (ARC-WFP) for their comments and feedback on the manuscript. Thanks also to UN Global Pulse, BBVA Data and Analytics and Bancomer teams for their support.